\documentclass[aps,nofootinbib]{revtex4-1}
\usepackage{epsfig}
\usepackage{float,amsmath,amstext,amsthm,amssymb,amsfonts}
\usepackage{graphics}
\usepackage{subfigure}

\makeatletter
\renewcommand*\@thesubfigure{(\alph{subfigure})\hskip\subfiglabelskip}
\makeatother

\begin{document}

\title{The importance of radiative scattering in heated heavy ion plasmas}

\author{J.~Fiedler}

\affiliation{{Lortzingstr.~51}, {50931~Cologne}, Germany}

\begin{abstract}
The transport of energy in heated plasmas requires the knowledge of
the radiation coefficients. These coefficients consist of
contribution of bremsstrahlung, photoionisation, bound-bound
transmissions and scattering. Scattering of photons on electrons is
taken into account by the model of Thomson, Klein-Nishina and the
first order angular momentum of Klein-Nishina. It is shown that
radiative scattering becomes an important part of energy transport in
high temperature plasmas. Moreover, the contribution of transport
correction to scattering is taken into account. The physics is
discussed on the example of a heated plutonium plasma at different
particle densities and temperatures in radiative equilibrium.
\end{abstract}

\maketitle

%\documentclass[doublecol,figures]{epl2} 
%\usepackage{tabularx}
%%\usepackage{graphics}
%\usepackage{amsmath}
%\usepackage{float}
%\usepackage{subfigure}
%
%%\renewcommand*\thesubfigure{-\arabic{subfigure}}
%\renewcommand*\thesubfigure{\alph{subfigure}}
%\makeatletter
%\renewcommand*\@thesubfigure{(\alph{subfigure})\hskip\subfiglabelskip}
%\makeatother
%
%\parindent=0pt
%
%\title{The importance of radiative scattering in heated heavy ion plasmas}
%
%\author{J. Fiedler}
%
%\institute{Lortzingstra\ss{}e 51, 50931 Cologne, Germany, EU}
%
%\pacs{44.40.+a}{Radiative transfer in heat transfer}
%\pacs{95.30.Jx}{Radiative transfer in astrophysics}
%
%\abstract{The transport of energy in heated plasmas requires the
%  knowledge of the radiation coefficients.  These coefficients consist
%  of contribution of bremsstrahlung, photoionisation, bound-bound
%  transmissions and scattering. Scattering of photons on electrons is
%  taken into account by the model of Thomson, Klein-Nishina and the
%  first order angular momentum of Klein-Nishina. It is shown that
%  radiative scattering becomes an important part of energy transport
%  in high temperature plasmas. Moreover, the contribution of transport
%  correction to scattering is taken into account. The physics is
%  discussed on the example of a heated plutonium plasma at different
%  particle densities and temperatures in radiative equilibrium.}
%
%\newcolumntype{Q}[1]{>{\centering\arraybackslash}p{#1}}
%\begin{document}
%
%\maketitle

\section{Introduction}

Calculating energy losses by radiative effects in heated plasmas
requires the knowledge of the radiation transport coefficients. In
local thermal equilibrium these coefficients strongly depend on the
electron density, the averaged ionisation stage and the ion density at
different stages of ionisation. These contributions are estimated by
the eq. of Saha \cite{Zel'dovich:66}. The required ionisation
potentials are taken from literature
\cite{Rodrigues:04}. Additionally, depression effects which lead to a
lowering of the ionisation potentials are considered. The radiation
transport coefficients are specific for each constituent of a plasma
and include absorption, emission and scattering contributions. It is
necessary to distinguish between the process of absorption and
scattering. A process is called \textit{scattering} when a photon
interacts with a scattering centre (e.g. atom, ion, electron) and
emerges from the interaction into a new direction without altered energy
(Thomson scattering) or a slightly altered energy (Compton
scattering). Scattering processes mainly depend upon the radiation
field and are only weakly coupled to the thermodynamic properties of
the plasma. A process is called absorption, when a photon is destroyed
by conversion of its energy (wholly or partly) into thermal energy of
the plasma \cite{Mihalas:78}. Sometimes those absorption processes are
called \textit{true} absorption. Absorption processes convert the
photon energy directly into thermal energy of the gas
\cite{Mihalas:78}. The physics of radiative scattering and absorption
is discussed in the example of a heated plutonium plasma in a range of
temperature between \mbox{10 eV} and \mbox{100 keV}. The problem of
radiative absorption is discussed widely in literature, e.g
\cite{Li:09}, whereas scattering is observed rarely, e.g.
\cite{Pritzker:76}. Within this report, the results for the overall
frequency integrated averaged mean free paths in radiative equilibrium
in the limit of dilute and dense plasmas taking scattering on
electrons at rest in transport correction into account are
presented. The knowledge of the frequency spectra integrated mean free
paths is of importance in the diffusion limit of the radiation
transport theory, e.g \cite{Olson:00}.

\section{Theory}

The required electron and ion densities in local thermal
equilibrium can be determined from the Saha
eq. \cite{Zel'dovich:66}. Assuming non-relativistic and
non-degenerated electrons, it is shown \cite{Fiedler:10} that this
eq. can be written in form of the non-linear eq.

\begin{equation}\label{eqn.rbarequation}
  \bar{r} + \sum_{j = 1}^{Z} \frac{1}{\bar{r}^j}\left(\bar{r} - j\right)
  \prod_{r=1}^j K^D_r = 0,
\end{equation}

where

\begin{equation}\label{eqn.KDr}
  K^D_r = \frac{2\left(2\pi m_e k_BT\right)^{3/2}}{h^3}
  \exp\left(\frac{-I_r + D}{k_BT}\right),
\end{equation}

$\bar{r} = n_e/n_p$ is the averaged stage of ionisation, $n_p$ is the
initial particle density, $n_e$ the electron density $k_B$ the
Boltzmann constant, $T$ the temperature, $I_r$ the ionisation
potential, $D$ the depression energy, $m_e$ the electron mass, $h$ the
Planck constant and $Z$ the atomic number. The participation
probabilities $n^*_r = n_r/n_p$ of the $r$-times ionised atoms are
obtained using the recursive relation

\begin{equation}\label{eqn.Saha}
  n^*_r = \frac{K^D_r}{n_p\;\bar{r}} n^*_{r-1},\qquad r = 0,\dots,Z,
\end{equation}

where $n_r$ and $n_{r-1}$ are the ion densities at the ionisation
levels $r$ and $r -1$, respectively. $n^*_0$ has to be determined
iteratively. The ionisation potentials are depressed by free electrons
in the vicinity of an ion. It is shown that the depression term $D$
can be written as \cite{Cox1:68}

\begin{equation}\label{eqn.Depression}
  D = 1.161\times 10^{-10}\bar{r}n^{1/3}_p\quad \mbox{(eV)}.
\end{equation}

$n_p$ is given in cm$^{-3}$. The radiative absorption contribution
consists of bound-bound, photo effect and (inverse) bremsstrahlung
contribution. The bound-bound input has been neglected in the current
case. The bremsstrahlung absorption contribution, also called
free-free absorption, reads \cite{Zel'dovich:66}

\begin{equation}
  \Sigma_{ff}(u) = 0.767\times 10^{-47}\frac{\bar{r}^3 n^2_p}{u^3k_BT}, 
  \quad (\mbox{cm}^{-1})
\end{equation}

where $u = h\nu/k_BT$ and $\nu$ is the photon frequency. For the case
of hydrogen-like ions the combined bremsstrahlung and photo effect
absorption cross section, also called bound-free absorption, can be
written as \cite{Zel'dovich:66}

\begin{equation}\label{eqn.GebundenFrei}
  \Sigma_{a}(u) = 0.96\times 10^{-7}\frac{n_p}{T^2} 
  \sum^{Z - 1}_{r = 0} n^*_r (r + 1)^2 F_r(u), \quad (\mbox{cm}^{-1})
\end{equation}

where $n_r^*$ is given by (\ref{eqn.Saha}). The function
$F_r(u)$ is given by

\begin{equation}
  F_r(u)= \left\{
    \begin{array}
      {r@{\quad:\quad}l}
      \displaystyle\frac{1}{u^3}\exp\left(u - \frac{I_{r + 1}}{k_BT}\right) & 
      \displaystyle u \le \frac{I_{r + 1}}{k_BT}\\[5mm]
      \displaystyle\frac{2}{u^3}\frac{I_{r + 1}}{k_BT} & 
      \displaystyle u > \frac{I_{r + 1}}{k_BT}.
    \end{array}
  \right.
\end{equation}

The temperature $T$ is measured in Kelvin. Due to the quantum
mechanical predictions of spontaneously and induced emitted photons,
the \textit{true} absorption contribution has to be corrected in the
following way, e.g. \cite{Armstrong:72}

\begin{equation}\label{eqn.EffAbsorption}
  \Sigma'_a = \Sigma_a\left(1-\exp(-u)\right).
\end{equation}

More precise data for the above given absorption coefficient in range
of photon energy \mbox{$h\nu = 100 - 2000$} eV are obtained by Henke
\textit{et~al.} \cite{Henke:84}. Besides absorption, photons can
interact with matter by scattering.  The probability that a photon
having frequency $\nu$ and flight direction $\mathbf{\Omega}$ is
scattered to frequency $\nu'$ and direction $\mathbf{\Omega'}$ within
the intervals $d\nu$ and $d\mathbf{\Omega}$ in travelling a distance
$ds$ is described by
\mbox{$\sigma_s(\nu\rightarrow\nu',\mathbf{\Omega}\cdot\mathbf{\Omega'})\;
d\nu\;d\mathbf{\Omega}\;ds$}, e.g. \cite{Pomraning:73}. The quantity
\mbox{$\sigma_s(\nu\rightarrow\nu',\mathbf{\Omega}\cdot\mathbf{\Omega'})$} is
called the differential scattering cross section. Within the
Klein-Nishina model this quantity reads \cite{Pomraning:73}

\begin{equation}\label{eqn.ComptonScattTransPritzker}
  \begin{split}
    \sigma_s(\nu\rightarrow\nu',\mu) &= 
    \frac{3}{16\pi}\frac{1}
    {\left(1 + \gamma(1 - \mu) \right)^3} 
    \bigg[1 + \gamma + \gamma^2
      - (\gamma + 2\gamma^2)\mu + (1 + \gamma + \gamma^2)\mu^2
      - \gamma\mu^3\bigg]\\
    &\times \delta\left(\nu' - \frac{\nu}{1 + \gamma(1 - \mu)}\right)
    \left(\frac{8\pi e^4}{3m^2_ec^2}\right),
  \end{split}
\end{equation}

where \mbox{$\gamma = h\nu/m_e c^2$} is a dimensionless energy, $h$ is
the Planck action, $\nu$ the frequency of the incoming photon, $\nu'$
the frequency of the photon after scattering, $c$ the speed of light,
$e$ is the electron charge and \mbox{$\mu =
\mathbf{\Omega}\cdot\mathbf{\Omega'}$}. Since one asks for the
scattering probability of photons with all possible frequencies $\nu'$
for all possible angles $\mu$ the differential scattering cross
section (\ref{eqn.ComptonScattTransPritzker}) is evaluated over all
frequencies and scattering directions. In that way one obtains the
macroscopic Klein-Nishina cross section

\begin{equation}\label{eqn.MacroscopicKleinNishinaCrossSection}
  \begin{split}
    \Sigma_s(\gamma) &= 2\pi n_e \int^1_{-1}d\mu\int^{\infty}_0d\nu'\;
    \sigma_s(\nu\rightarrow\nu',\mu)\\
    &= \frac{3}{4}n_e\left[\frac{1 + \gamma}{\gamma^3}
      \left(\frac{2\gamma\left[1 +\gamma\right]}
      {1 + 2\gamma} - \log\left(1 + 2\gamma\right)\right)
      + \frac{1}{2\gamma}\log\left(1 + 2\gamma\right) - 
      \frac{1 + 3\gamma}{\left(1 + 2\gamma\right)^2} \right]
    \left(\frac{8\pi e^4}{3m^2_ec^2}\right).
  \end{split}
\end{equation}

The Klein-Nishina scattering contribution reaches a maximum at around
\mbox{5 keV} for a plutonium plasma at normal particle density. Refer
to fig. (\ref{fig.1}). By increasing temperature the scattering
absorption decreases. This is due to the upcoming pair-production at
very high temperatures. For later purpose the first angular momentum of
(\ref{eqn.ComptonScattTransPritzker}) is introduced. The multiplication
of (\ref{eqn.ComptonScattTransPritzker}) by $\mu$ followed by an
integration over all photon frequencies and scattering directions leads
to 

\begin{equation}
  \begin{split}
    A_1(\gamma) &= 2\pi n_e \int^1_{-1}d\mu \int^{\infty}_0d\nu'\;
    \mu\sigma_s(\nu\rightarrow\nu',\mu)\\
    &= \frac{3}{4}\frac{n_e}{\gamma^4(1 + 2\gamma)^2}
    \bigg[\Big(4\gamma^5 - 27\gamma^3 - 37\gamma^2
      - 18\gamma - 3\Big)\log\left(1 + 2\gamma\right)
      - 6\gamma^5 + 16\gamma^4\\
      &{} + 46\gamma^3 + 30\gamma^2 + 6\gamma\bigg]
    \left(\frac{8\pi e^4}{3m^2_ec^2}\right).
  \end{split}
\end{equation}

Neglecting the dependency of scattering on $\gamma$ in
(\ref{eqn.MacroscopicKleinNishinaCrossSection}) the scattering of
photons on free electrons at rest is described by the macroscopic
Thomson coefficient \cite{Pomraning:73}

\begin{equation}\label{eqn.MacroscopicThomsonCrossSection}
  \Sigma_{Th} = \sigma_{Th} n_e = \frac{8\pi e^4}{3m^2_e c^2} n_e,
\end{equation}

where $\sigma_{Th}$ is the frequency independent microscopic Thomson
coefficient. $\Sigma_{Th}$ is valid in very low photon energy limits
only. Scattering does not appear spontaneously. Hence, the scattering
contribution must not be corrected similar to
(\ref{eqn.EffAbsorption}). Quantum-mechanical corrections applied to
the absorption coefficient using the Gaunt factor \cite{Pomraning:73}
are of very small effect only and have not been considered here.
Often, it is of practical interest having radiation constants
depending on temperature and material density only. For such cases
photon frequency averaged methods have been developed. Of importance
are the one-group photon energy coefficient. These coefficients are
integrated over all photon energy frequencies. For the cases of optical
thin and thick plasmas in radiative equilibrium the weight functions
of Planck and Rosseland are widely-used, e.g. \cite{Tsakiris:87}. In
an optical thick regime the photon is destroyed in the vicinity of its
emergence. The averaged mean free path used in this environment is
called the Rosseland mean free path $\lambda_R$ and is defined as
\cite{Zel'dovich:66}

\begin{equation}\label{eqn.RosselandMeanFreePath}
  \lambda_{R} = \Sigma^{-1}_R = \frac{15}{4\pi^4}\int^{\infty}_0 \frac{du}
         {\Sigma_{tr}} \frac{u^4\exp(-u)}{\left(1 - \exp(-u)\right)^{-2}},
\end{equation}

where $\Sigma_{tr}$ is the transport cross section. $\Sigma_R$ is the
Rosseland mean absorption coefficient. A regime is called optical thin
when the averaged mean free path is in range or above of the
dimensions of the underlying physical system. In contrast to
(\ref{eqn.RosselandMeanFreePath}) the transport cross section has to
be averaged by the weighting function of Planck.  In that case the
averaged mean free path $\lambda_P$ is given by \cite{Zel'dovich:66}

\begin{equation}\label{eqn.PlanckMeanFreePath}
  \lambda_P = \Sigma^{-1}_P = \frac{\pi^4}{15}\left(\int^{\infty}_0 du\;
  \Sigma_{tr}\frac{u^3\exp(-u)}{1 - \exp(-u)}\right)^{-1}.
\end{equation}

$\Sigma_P$ is the Planck mean absorption coefficient. The extension of
the formulae (\ref{eqn.RosselandMeanFreePath}) and
(\ref{eqn.PlanckMeanFreePath}) to a multi-group approach is
straight-forward. The transport cross section is defined by
\cite{Pomraning:73}

\begin{equation}
  \Sigma_{tr} = \Sigma'_a + \Sigma_s - A_1.
\end{equation}

$A_1$ vanishes when scattering is taken into account in the Thomson limit. In
that case the transport cross section is equal to the total cross
section

\begin{equation}
  \Sigma_{tr} = \Sigma_{tot} = \Sigma'_a + \Sigma_s.
\end{equation}

Bernstein \cite{Bernstein:03} proved mathematically that $\Sigma_P$
forms an upper limit for $\Sigma_R$ to within a factor of close to
unity. For scattering this relation is shown in
\mbox{fig. (\ref{fig.1})}.

\section{Results}

\begin{figure}
  \includegraphics[width=88mm]{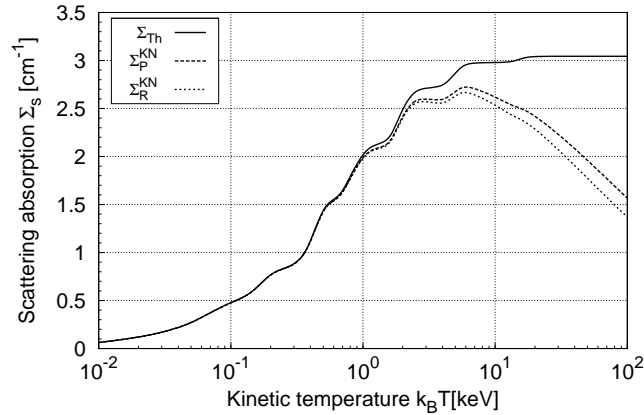}
  \caption{Mean averaged scattering absorption contribution for
    plutonium at normal particle density. $\Sigma_{Th}$ means the
    Thomson scattering, $\Sigma^{KN}_{P}$ is the Planck averaged and
    $\Sigma^{KN}_{R}$ the Rosseland averaged Klein-Nishina absorption.}
  \label{fig.1}
\end{figure}

\begin{figure*}
  \begin{center}
    \subfigure[ $n_p$ = 4.87$\times 10^{20}$ cm$^{-3}$.\label{fig.2a}]
              {\includegraphics[width=0.49\textwidth]{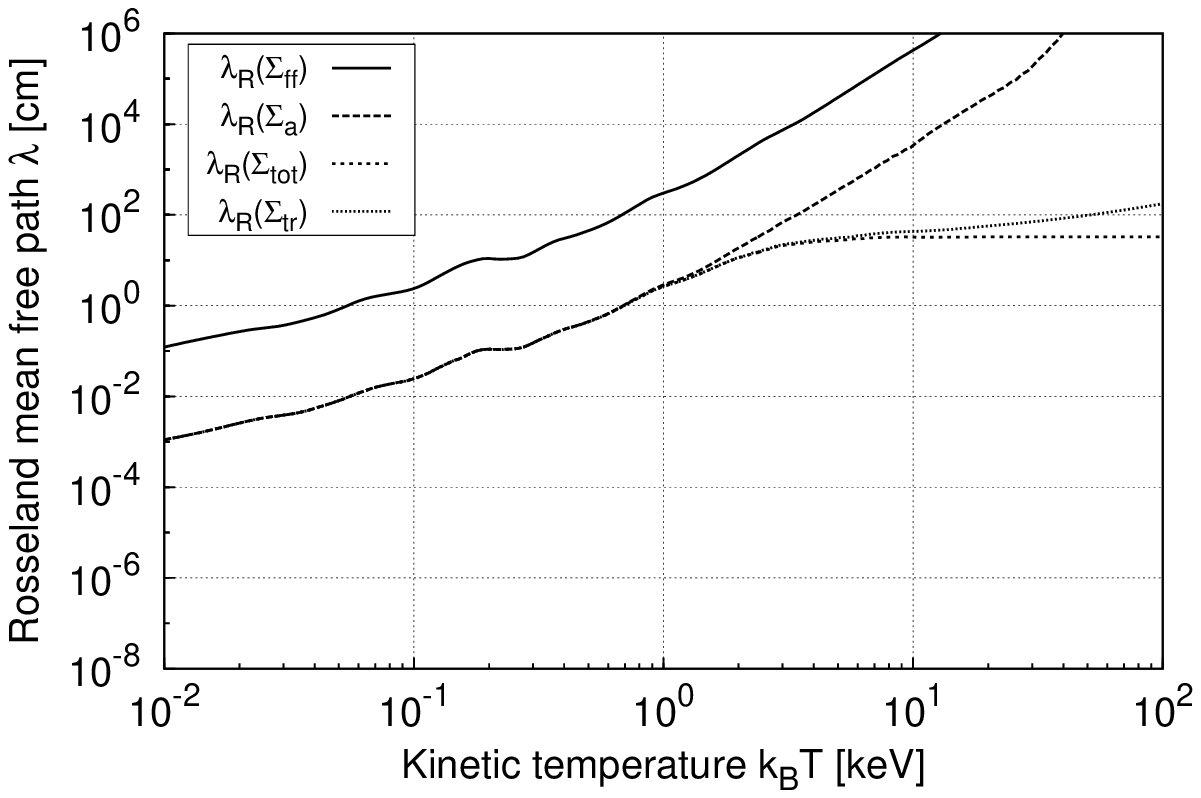}}\hfill
    \subfigure[ $n_p$ = 4.87$\times 10^{20}$ cm$^{-3}$.\label{fig.2b}]
              {\includegraphics[width=0.49\textwidth]{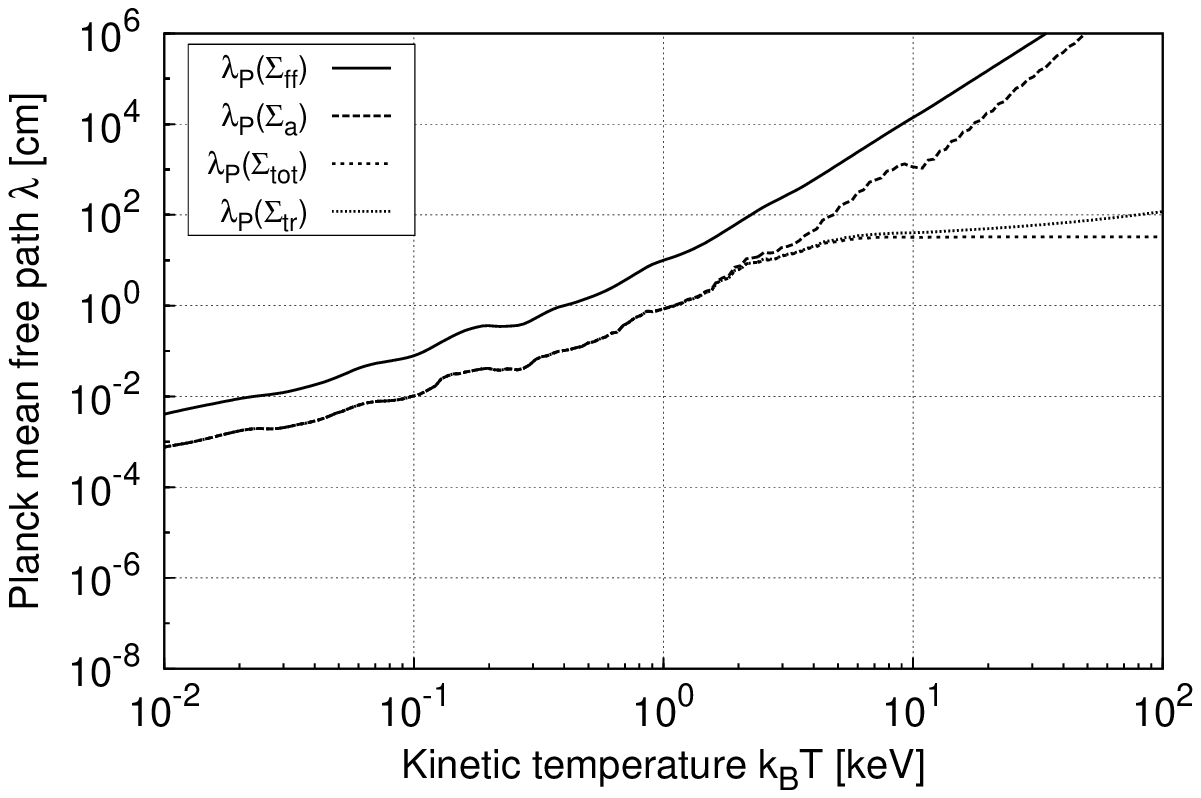}}
    \subfigure[ $n_p$ = 4.87$\times 10^{22}$ cm$^{-3}$.\label{fig.2c}]
              {\includegraphics[width=0.49\textwidth]{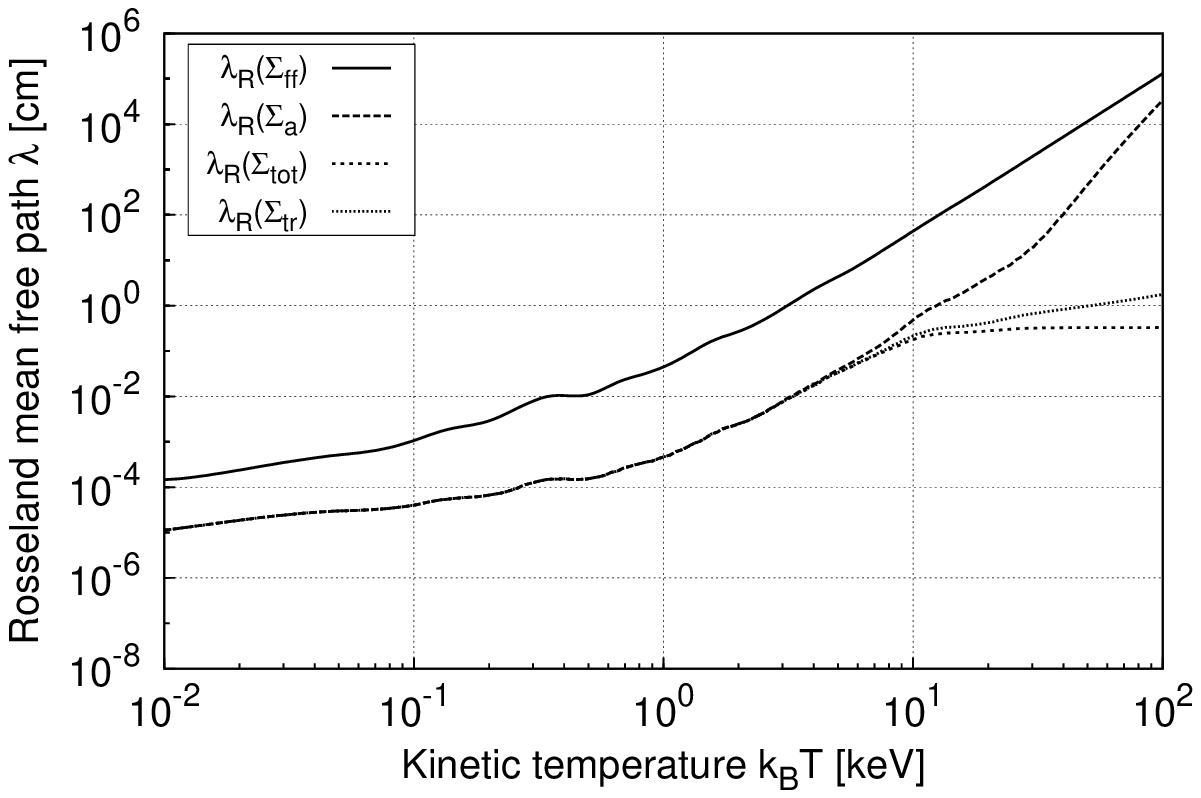}}\hfill
    \subfigure[ $n_p$ = 4.87$\times 10^{22}$ cm$^{-3}$.\label{fig.2d}]
              {\includegraphics[width=0.49\textwidth]{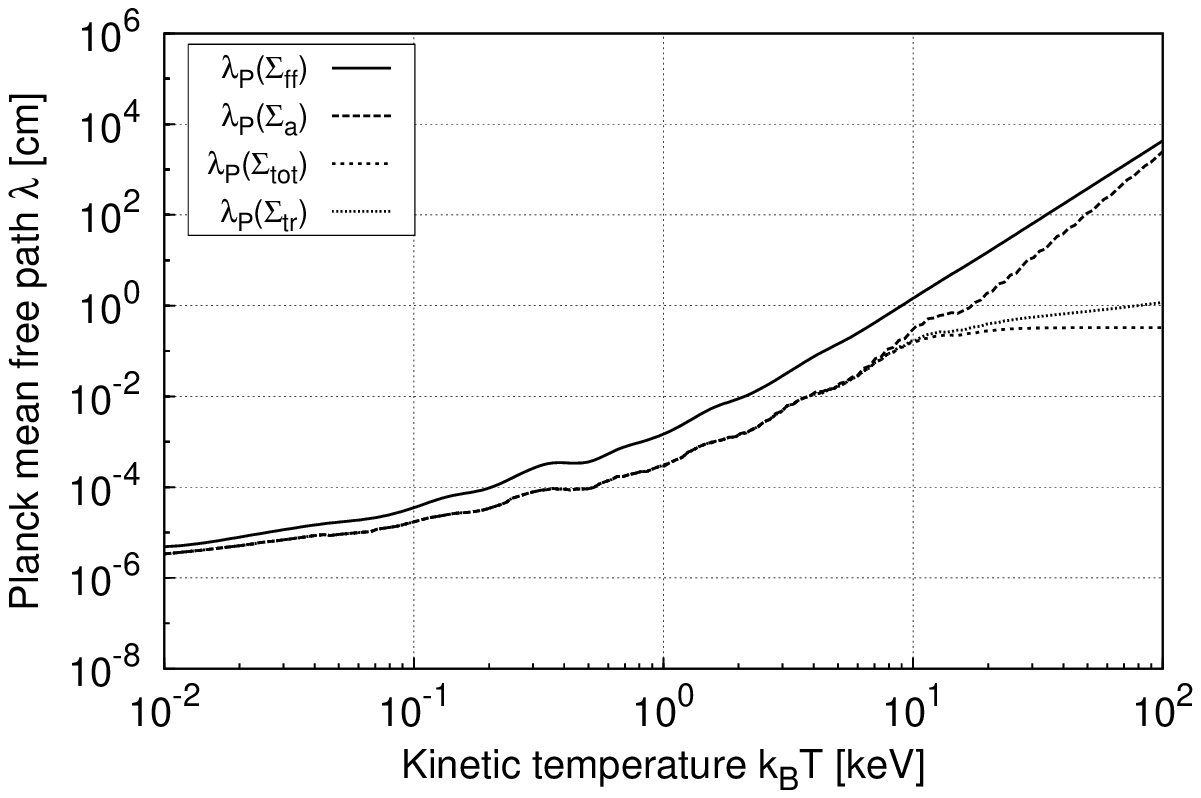}}
    \subfigure[ $n_p$ = 4.87$\times 10^{24}$ cm$^{-3}$.\label{fig.2e}]
              {\includegraphics[width=0.49\textwidth]{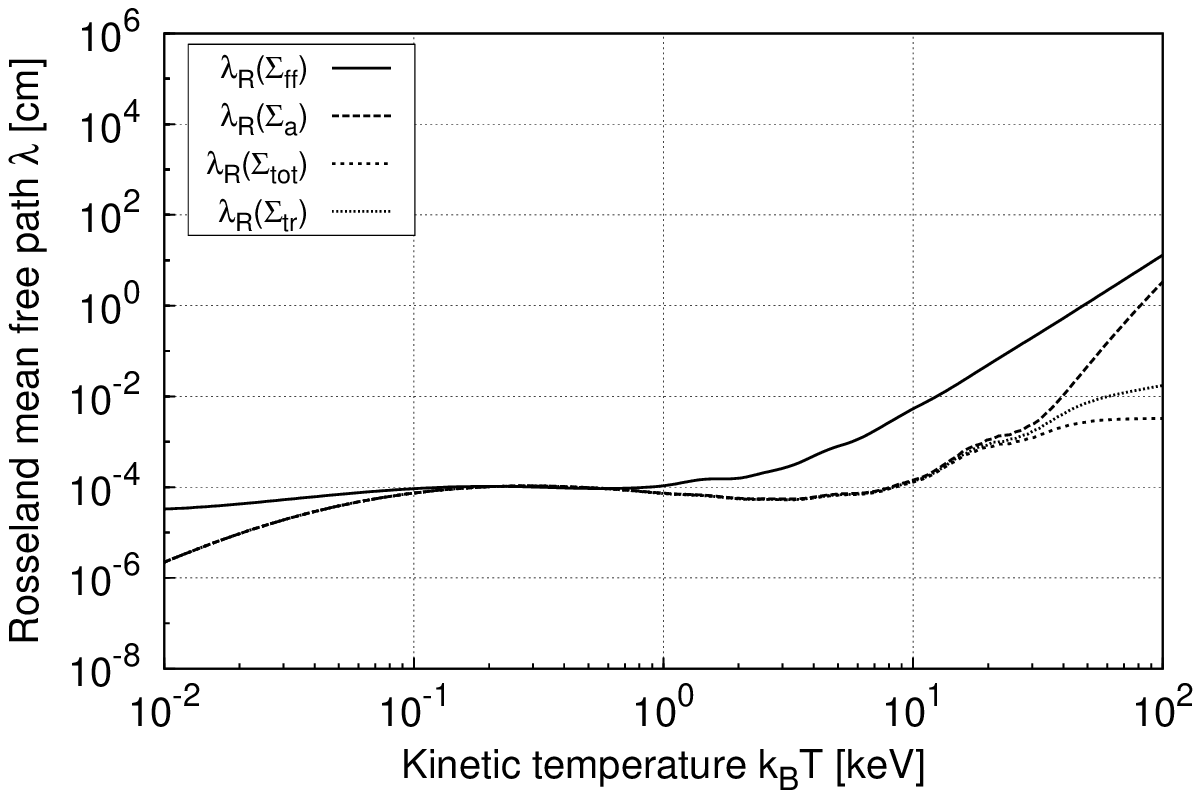}}\hfill
    \subfigure[ $n_p$ = 4.87$\times 10^{24}$ cm$^{-3}$.\label{fig.2f}]
              {\includegraphics[width=0.49\textwidth]{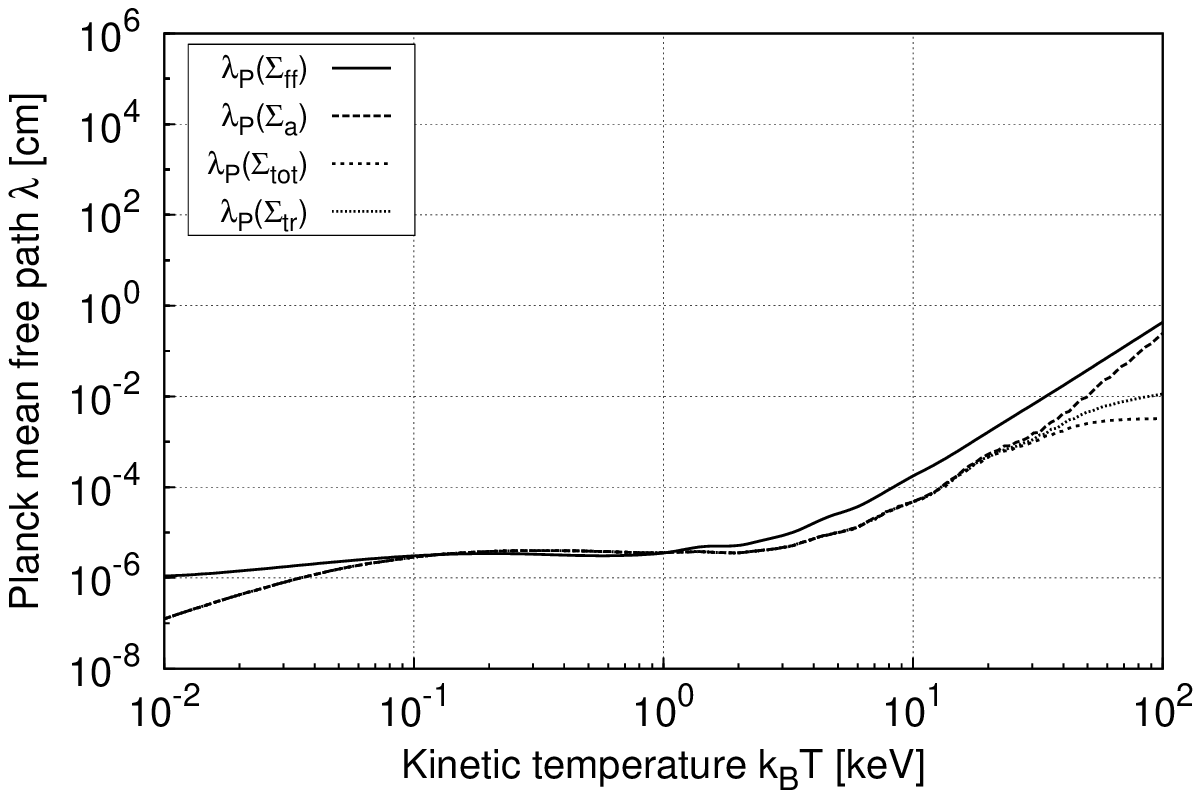}}
    \caption{Averaged mean free paths for the bremsstrahlung,
      $\Sigma_{ff}$, the combined bremsstrahlung and photo-effect,
      $\Sigma_a$, the total cross section, $\Sigma_{tot}$, and the
      transport cross section, $\Sigma_{tr}$, in a plutonium plasma
      at different particle densities depending on temperature.}
    \label{fig.2}
  \end{center}
\end{figure*}

The results presented in the last section are applied to a heated
plutonium plasma at different densities. Under normal circumstances,
the particle density of plutonium ($\alpha$-phase) is \mbox{$n_p =
  4.87\times 10^{22}$ cm$^{-3}$}. \mbox{Figure (\ref{fig.1})} shows
the influence of the scattering contribution of the models of Thomson
and Klein-Nishina depending on temperature for a plutonium plasma at
normal particle density. At high temperature the scattering absorption
by the model of Thomson saturates to a constant due to the fully
ionisation of the plasma. The wave-like structure is a result of the
ionisation behaviour of plutonium \cite{Fiedler:10}. Figure
(\ref{fig.2}) shows the Planck and Rosseland averaged mean free path
of the pure bremsstrahlung $\Sigma_{ff}$, the combined photo-effect
and bremsstrahlung contribution $\Sigma_a$, the total cross section
$\Sigma_{tot}$ and the transport cross section $\Sigma_{tr}$. The
calculations have been performed with a density of one percent of the
normal particle density, fig. (\ref{fig.2a}) and (\ref{fig.2b}), at
solid state density, fig. (\ref{fig.2c}) and (\ref{fig.2d}) and
100-times compression compared to the solid state density,
fig. (\ref{fig.2e}) and (\ref{fig.2f}). In a dilute plasma the free
electrons are weakly influenced by the electric field of the
surrounding ions. In contrast to an electron being in the field of an
ion, the photon energy absorbed by an electron which does not belong
to the vicinity of an ion is not converted into thermal energy. The
lower the ion density of the plasma the more important scattering
becomes. This effect is shown in fig. (\ref{fig.2}). The Thomson limit
is valid in a small range of temperature only. At a temperature of
approximately \mbox{1 keV} the Klein-Nishina scattering becomes
important. Refer to fig. (\ref{fig.1}). Especially in thin plasmas
scattering significantly changes the absorption behaviour of radiation
and cannot be neglected. Furthermore, \mbox{fig. (\ref{fig.2})} shows
the relevance of absorption by the photo-effect at mid-ranged
temperatures. This contribution eases at very high temperatures when
the plasma becomes fully ionised.

\acknowledgments I am grateful to R. {\sc K\"ulheim} and A. {\sc
  Gro\ss{}mann} giving corrections and comments.

\vfill\eject

\end{document}